\documentstyle[psfig]{article}
\begin{document}
% Zeugs fuer die Bildli
%\input epsf         % defines \epsfbox and supporting macros
%\epsfverbosetrue    % messages will show height and width
% scale down pictures if they don't fit into page 
%\def\epsfsize#1#2{\ifnum#1>\hsize\hsize\else#1\fi} 

\newcommand{\vp}{\varphi}
\newcommand{\ra}{\rightarrow}
\newtheorem{theorem}{Theorem}[section]
\newtheorem{proposition}{Proposition}[section]
\newtheorem{lemma}{Lemma}[section]
\newtheorem{corollary}{Corollary}[section]
\newtheorem{conjecture}{Conjecture}[section]
\newtheorem{example}{Example}[section]
\begin{center}
{\LARGE \bf Are There Static Textures?}\vspace{1cm}\\
{\bf Lukas Lichtensteiger$^1$ and Ruth Durrer$^2$}\\
$^1$AILab, Computer Science Department, Universit\"at Z\"urich, 
Winterthurerstrasse 190, CH-8057 Z\"urich, Switzerland\\
$^2$D\'epartement de Physique Th\'eorique, Universit\'e de Gen\`eve,
Quai Ernest Ansermet 24, CH-1211 Gen\`eve, Switzerland
\end{center}
\vspace{1.5cm}
\begin{abstract}
We consider harmonic maps from Minkowski space into the three sphere.
We are especially interested in solutions which are asymptotically
constant, i.e. converge to the same value in all directions of spatial
infinity. Physical 3-space can then be compactified and can be
identified topologically 
(but not metrically!)  with a three sphere. Therefore, at fixed
time, the winding of the map is defined. We investigate whether static
solutions with non-trivial winding number exist. The answer which we
can proof here is only partial: We show that within a certain family
of maps no static solutions with non-zero winding number exist. We
discuss the existing static solutions in our family of maps. An
extension to other maps or a proof that our family of maps is
sufficiently general remains an open problem. 
\end{abstract}

\section{Introduction}
Many physical problems can be described by scalar fields $\vp$
with topologically non-trivial target spaces. The equation of motion
for $\vp$ often requires $\vp$ to represent a harmonic map from
spacetime into the target space (In the physics literature such
maps are better known under the name 'non-linear $\sigma$-model').

The question arises
whether a given field configuration is topologically trivial 
(continuously deformable to the constant map). 
Topological defects are topologically
non-trivial field configurations. If we consider a topologically trivial 
four dimensional space-time manifold $\cal M$, non-trivial field 
configurations  are in general singular on a certain sub-manifold 
${\cal S}\subset {\cal M}$. The dimension of $\cal S$ depends on the first
non-trivial homotopy of $\vp({\cal M}\backslash{\cal S})\equiv Im\vp$: \\
If 
$\pi_0(Im\vp)$ is non-trivial, the submanifold $\cal S$ forms a
network of 'domain walls' of space-time dimension 3.
If $\pi_1(Im\vp)$ is non-trivial, a network of 'strings' of space-time
dimension 2 is formed. If 
$\pi_2(Im\vp)$ is non-trivial 'monopoles' of space-time dimension 1
appear. And if 
$\pi_3(Im\vp)$ is non-trivial 'textures', singular events of 
space-time dimension 0 appear.

Higher homotopy groups do not lead to topological defects in four
space-time dimensions.
The simplest and most common examples of topological defects are the
cases $Im\vp =S^n$, where $S^n$ denotes the sphere of dimension
$n$ and $S^0=\{-1,1\}$.
But also other examples play an important role in solid state physics
(helium\cite{KiVo}, liquid crystals\cite{CDTY}) and cosmology\cite{Ki,Ruth}.

In the case of a field living on $S^n$ with $n\le2$ simple static
domain wall~($n=0$), string~($n=1$) and monopole~($n=2$) solutions 
are known (see,
e.g. \cite{Ruth} or \cite{VS}). The question which we want to address
here is whether there also exist static texture solutions ($n=3$).
A static texture solution is different from monopoles, strings and
domain walls in that it is non-singular: A map from $\bf R^3$ to $S^3$
which is asymptotically constant, $\lim_{|x|\ra\infty}\vp(x,t)=\vp_0(t)$, can
wind around $S^3$ without being singular anywhere. Derricks theorem\cite{De}
then already implies that there is no static texture solution with
finite energy. However, the simple static domain wall, string and monopole
solutions we are alluding to (which contain a singular sheet, line
and point respectively), have infinite total energy and we thus
want to allow also for infinite energy solutions. We therefore cannot
apply Derricks theorem. Nevertheless, numerical
simulations\cite{DHZ,DZ} indicate, that
winding texture configurations always shrink, leading to a
singularity, the unwinding event, at a finite time $t_c$, after which
the configuration becomes topologically trivial and approaches the
constant solution, as predicted by Derricks theorem. The total energy
of the initial configuration is, however in general infinite so that
Derricks theorem cannot be applied. 

This numerical finding prompted us to search for a proof for the
non-existence of static texture in flat physical
space. 
Clearly, the result depends on the geometry of physical space. If
 space is a 3-sphere, the identity map represents a well
defined static texture solution. We want to investigate whether such
solutions are excluded, for example in Minkowski space.

We do not quite succeed in this task. First, we shall assume that 
the searched for static winding solution has
a spherically symmetric Lagrangian density. This assumption does not
bother us too much. It seems physically well motivated (we can,
however not use any rigorous energy arguments to justify it, since the
total energy of our solution must be infinite).  Also
within the class of solutions with spherically symmetric Lagrangian
densities we have a proof only for a special ansatz for the
field configuration and it remains an open problem how general our
ansatz is.

Our paper is organized as follows: In the next section we write down
the equations of motion for the scalar field and specify our ansatz.
In section~3 we then show that within this ansatz no static solution
with non-trivial winding number can exist and discuss the nature of
the globally existing (non-winding) static solutions. In section~4 we 
present the conclusions and an outlook.

\section{Spherically symmetric 'texture' fields}
We consider a scalar field (order parameter)  
$\varphi:{\cal M}\rightarrow S^{3}$ and we ask $\vp$ to be harmonic, a
non-linear $\sigma$-model. A harmonic map satisfies the 
Euler-Lagrange equations for the action 
\begin{equation}
	{\cal S}(\varphi)=\frac{1}{2}\int_{\cal M}|d\varphi|_{S^{3}}^{2}dx.
	\label{Lagrangian}
\end{equation}
We consider the situation where $\cal M$ is 4-dimensional Minkowski 
spacetime with the flat Lorentzian metric $g$ and
$S^{3}$ is the unit three-sphere with the standard metric which we 
denote by $G$.

Here $dx$ denotes the volume element of the metric $g$ on 
$\cal M$ and 
\begin{equation}
{\cal L}(\vp)=	  |d\varphi|_{S^{3}}^{2}
	  = \mbox{Trace}_{g}(\varphi^{*}G)
	  = g^{\mu\nu}(x)G_{ij}(\varphi(x))
	  \frac{\partial\xi^{i}}{\partial x_{\mu}}(x)
	  \frac{\partial\xi^{j}}{\partial x_{\nu}}(x),	
	\label{Df}
\end{equation}
for some (local) coordinates $x=(x_{0},\ldots,x_{3})$ on $\cal M$ and 
$\varphi=(\xi_{1},\xi_2,\xi_{3})$ on $S^{3}$ (We always assume summation 
over repeated indices). 

 We only consider regular maps $\varphi$, i.e. maps that have
finite energy density $|d\varphi|_{S^{3}}^{2}$ everywhere.
In addition to being stationary points of (\ref{Lagrangian}) we also 
demand  our maps to be asymptotically constant,
\[ \lim_{|x|\ra\infty}\vp(x,t)= \vp_0(t) ~. \]
At fixed time we then can consider them as maps $\overline{\vp}_t$
from compactified ${\bf R}^3$, $\overline{{ \bf R}^3} = 
{\bf R}^3\cup\{\infty\} 
\equiv S^3$ to $S^3$, assigning $\overline{\vp}_t(x)=\vp(x,t)$ and 
$\overline{\vp}_t(\infty,t)=\vp_0(t)$. 

The winding number of this extended map $\overline{\vp}_t :S^3\ra S^3$ 
is a topological invariant and counts how many times $\overline{\vp}_t(S^3)$
winds around the target $S^3$. This number cannot change under
continuous time evolution.
We would like to show that there are no static solutions $\vp$ with
non-zero winding number.

Unfortunately, we are not able to solve the problem in this
generality. We thus
impose some restrictions on the maps $\varphi$. One way of 
doing this is to demand $\vp$ to obey certain symmetry properties.
We want to impose  spherical symmetry, i.e., 
invariance under $SO(3)$, the group of rotations of physical space. 

The action of an element of the rotation group, $g\in SO(3)$ on the maps 
$\varphi:M\rightarrow S^{3}$ is given by
\begin{equation}
	(g\cdot\varphi)(x)=\varphi(g^{-1}\cdot x)
	\label{GroupAction2}
\end{equation}
(scalar field). The fixed points of the action 
(\ref{GroupAction2}) are the spherically symmetric fields of the 
form $\varphi=\varphi(r,t),\;r=|\vec{x}|$. We might want to require
spherical symmetry of the field $\vp$ itself. For our purposes however, this 
restriction is too severe: Since the image of a smooth map can never have a
dimension greater than the dimension of the domain, this would limit us to 
only two-dimensional ranges (one-dimensional in the static case) 
which are topologically not interesting. Instead we will only demand that 
the Lagrangian density 
\begin{equation}
{\cal L}=|d\varphi|_{S^{3}}^{2} \label{4}
\end{equation}
be $SO(3)$ invariant, ${\cal L}(g\varphi)={\cal L}(\varphi)$. 

We proceed as follows: We first derive the full 
Euler-Lagrange equations and then make an ansatz for $\varphi$ 
(which is inspired by our symmetry requirement).  We then 
insert the ansatz into the Euler-Lagrange equations. Since
 the equations remain 
self-consistent, we can try and solve them. The solutions we find are 
then always solutions of the full Euler-Lagrange equations. It remains 
 to investigate whether they can be topologically nontrivial, 
i.e., whether there exist solutions with nonzero winding number.

We use standard spherical coordinates $(r,\theta,\phi)$ for the spatial 
part and  write the standard metric on flat Minkowski spacetime $M$ as
\begin{equation}
	g=-dt^{2}+dr^{2}+r^{2}(d\theta^{2}+\sin^{2}(\theta)d\phi^{2}).
\end{equation}
For $y\in S^{3}$ we use the standard spherical coordinates
\begin{eqnarray}
	\nonumber
	y &=& (\sin\xi_{3}\sin\xi_{2}\sin\xi_{1},
	\sin\xi_{3}\sin\xi_{2}\cos\xi_{1},
	\sin\xi_{3}\cos\xi_{2},\cos\xi_{3}).
	\label{StandardKoord}
\end{eqnarray}
In our case the $\xi_{i}$ are functions living on 
spacetime $M$, 
\begin{equation}
    \xi_{i}:M\rightarrow {\bf R},\; 
    \xi_{i}=\xi_{i}(t,r,\theta,\phi).
\end{equation}
The standard ranges for the angles $\xi_i$ are $\xi_1\in [0,2\pi]$,
$\xi_2\in[0,\pi]$ and $\xi_3\in[0,\pi]$. It is important to note that
for a map to cover all of $S^3$, $\xi_2$ and $\xi_3$ have to assume
both boundary values, $0$ and $\pi$, and $\xi_{1}$ must assume both
$0$ and $2\pi$.

The standard metric on $S^{3}$ expressed in terms of $(\xi_i)$ is
\begin{equation}
	G = d\xi_{3}^{2} + d\xi_{2}^{2}\sin^{2}(\xi_{3})
	+ d\xi_{1}^{2}\sin^{2}(\xi_{3})\sin^{2}(\xi_{2}).
	\label{StandardM}
\end{equation}

The equations of motion corresponding to the Lagrangian 
(\ref{Lagrangian}) are
\begin{equation}
	\nabla_{\mu}\left(\frac{\partial{\cal L}}{\partial(\nabla_{\mu}\xi_{i})}
	\right)=\frac{\partial{\cal L}}{\partial\xi_{i}}, \;\;1\leq i\leq 3.
\end{equation}
With the standard metric (\ref{StandardM}) on $S^{3}$ they become:
%\begin{displaymath}
%	(1\leq i\leq 3):	
%\end{displaymath}
%\begin{eqnarray}
%	\nonumber
%	0&=&\nabla^{\mu}\nabla_{\mu}\xi_{i}+2\sum_{j=i+1}^{3}\cot(\xi_{j})
%	(\nabla_{\mu}\xi_{j})(\nabla^{\mu}\xi_{i})\\
%	&-&\cot(\xi_{i})\sum_{j=1}^{i-1}\sin^{2}(\xi_{i})\cdots\sin^{2}(\xi_{j+1})
%	(\nabla_{\mu}\xi_{j})(\nabla^{\mu}\xi_{j}).
%	\label{ELStd}
%\end{eqnarray}
%(The sum $\sum_{j=i+1}^{3}$ is to be omitted if $i=3$.
\begin{eqnarray}
    \nonumber
    0&=&\nabla^{\mu}\nabla_{\mu}\xi_{3}-\sin(\xi_{3})\cos(\xi_{3})
    \left[ (\nabla_{\mu}\xi_{2})(\nabla^{\mu}\xi_{2})+\sin^{2}(\xi_{2})
    (\nabla_{\mu}\xi_{1})(\nabla^{\mu}\xi_{1})\right]\\
    0&=&\nabla^{\mu}\nabla_{\mu}\xi_{2}+2\cot(\xi_{3})
    (\nabla_{\mu}\xi_{3})(\nabla^{\mu}\xi_{2})-\sin(\xi_{2})\cos(\xi_{2})
    (\nabla_{\mu}\xi_{1})(\nabla^{\mu}\xi_{1})\\
    \nonumber
    0&=&\nabla^{\mu}\nabla_{\mu}\xi_{1}+2\cot(\xi_{2})
    (\nabla_{\mu}\xi_{2})(\nabla^{\mu}\xi_{1})+2\cot(\xi_{3})
    (\nabla_{\mu}\xi_{3})(\nabla^{\mu}\xi_{1})
	\label{ELStd}
\end{eqnarray}

To obtain a spherically symmetric Lagrangian density we use a 
generalized hedgehog Ansatz:
\begin{equation}
	\xi_{i}=\xi_{i}(\phi,\theta), \;i=1,2\mbox{ and }
	\xi_{3}=\xi_{3}(r,t).
	\label{Ansatz}
\end{equation}
The idea behind this Ansatz is that  we want to make use on the vaste
knowledge on harmonic maps on 
2-dimensional spaces, which will help us to first solve the 
2-dimensional problem for $\xi_{1}$ and $\xi_{2}$ in the coordinates 
$\theta$ and $\phi$ and then afterwards solve the equation for the 
remaining function $\xi_{3}$. The crucial 
point in order for this to work is that the equations of motion
respects this `decomposition' of $\varphi$ in a $(r,t)$-dependent 
and a $(\theta,\phi)$-dependent part. This is the subject of the theory 
of $(\rho,\sigma)$-equivariant maps which is explained in great detail in 
\cite{Eells}. 

With the Ansatz (\ref{Ansatz}) we may introduce a map
\begin{eqnarray}
	\Omega: S^{2}\rightarrow S^{2}& , &\Omega(\phi,\theta)=(\xi_{1},\xi_{2})
	\label{Omega}
\end{eqnarray}
with Lagrangian density
\begin{equation}
	|d\Omega|^{2} = (\nabla_{\mu}\xi_{2})(\nabla^{\mu}\xi_{2})
	+\sin^{2}(\xi_{2})(\nabla_{\mu}\xi_{1})(\nabla^{\mu}\xi_{1}).
\end{equation}
The total Lagrangian density of the map $\varphi$ is then
\begin{equation}
	|d\varphi|^{2}=(\nabla_{\mu}\xi_{3})(\nabla^{\mu}\xi_{3})
	+\sin^{2}(\xi_{3})|d\Omega|^{2}.
	\label{EnergyStd}
\end{equation}
Our Ansatz (\ref{Ansatz}) yields
\begin{equation}
	|d\Omega|^{2}=\frac{\lambda(\phi,\theta)}{r^{2}},
	\label{ConstDO}
\end{equation}
Spherical symmetry of the Lagrangian density then requires 
$\lambda=constant (\geq 0)$. Inserting our Ansatz 
(\ref{Ansatz}) into the Euler-Lagrange equations (\ref{ELStd}),
we find after multiplying with $r^{2}$ that for the components 
$\xi_{1}$ and $\xi_{2}$ these equations are just 
the Euler-Lagrange equations of the map $\Omega:S^{2}\rightarrow S^{2}$. 
Thus $\Omega$ has to be harmonic (on $S^{2}$!) with constant Lagrangian density 
$|d\Omega|_{S^{2}}^{2}=\lambda$, where $|.|_{S^{2}}^{2}$ now denotes the 
Lagrangian density on $S^{2}$. But this means that $\Omega$ has to be an 
eigenmap of the Laplacian $\triangle$ on $S^{2}$ with eigenvalue 
$\lambda$ (in the sense of \cite{Eells}). Therefore the
components $\Omega_i$ have to be given by linear combinations of 
spherical harmonics of fixed degree $k$, $Y_{km}$,
\[ \Omega_i = \sum_m a_{im}Y_{km} ~.\] 

If we apply this to our map $\Omega:S^{2}\rightarrow S^{2}$, we obtain 
for $|d\Omega|^{2}$ (with respect to the metric $g$ on spacetime)
\begin{equation}
	|d\Omega|^{2}=\frac{k(k+1)}{r^{2}},\;k\in {\bf N}.
	\label{DO}
\end{equation}
We will always assume $\lambda=k(k+1)>0$ since we are only interested in 
spherically non-trivial maps. Note that $\lambda=2$ just corresponds to 
$\Omega=id$, the identity map, used for example in the `hedgehog' 
monopole.  

The remaining Euler-Lagrange equation for the last component $\xi_{3}$ 
is now
\begin{equation}
	0=\nabla^{\mu}\nabla_{\mu}\xi_{3} 
	-\frac{k(k+1)}{2r^{2}}\sin(2\xi_{3}),\;k\in {\bf N}.
	\label{EL3}
\end{equation}

It is this equation that we would like to analyze in this paper. In 
the next section we will prove the non-existence of \emph{static} 
solutions with nonzero winding number, whereas an infinite family 
of \emph{time-dependent} solutions of (\ref{EL3}) in Minkowski space
 for arbitrary winding number 
$n\in {\bf N}$ can be found \cite{LukasSpherical}.

In \cite{Bizon} solutions from a geometrical $S^3$ to $S^3$ are
studied with the ansatz (\ref{EL3}) for $k=0,1$. Here we consider
compactified ${\bf R}^3$, which is topologically equivalent to $S^3$, but
with flat geometry. 
Using our ansatz (\ref{Ansatz}) on the geometrical $S^{3}$,
we can also generalize the results 
of \cite{Bizon} by showing that there 
are actually two countable families of such maps for every 
$k>0,\;k\in {\bf N}$, where $k(k+1)$ is the eigenvalue of the map 
$\Omega$ defined in (\ref{Omega}). This will be detailed in a subsequent 
paper \cite{LukasSpherical}.

All of these results can also be found in \cite{Diplomarbeit}.

%%%%%%%%%%%%%%%%%%%%%%%%%%%%%%%%%%%%%%%%%%%%%%%%%%%%%%%%%%%%%%%%%%%%%%%%%

\section{Non-existence of static winding solutions}
\label{NostaticSect}
 We consider maps to the standard three sphere where 
spacetime is parametrized by standard Cartesian coordinates $r$ and $t$.
In what follows we will  use the notation
\begin{equation} 
     \dot{ }=\frac{\partial}{\partial t},\mbox{ and } 
     ^{\prime}=\frac{\partial}{\partial r}.
\end{equation}

Then (\ref{EL3}) becomes
\begin{equation}
	\xi_{3}^{\prime\prime}-\ddot{\xi_{3}}+\frac{2}{r}\xi_{3}^{\prime}
	-\frac{k(k+1)}{2r^{2}}\sin(2\xi_{3})=0,\;k\in {\bf N}
	\label{Glg1all}
\end{equation}
An exact solution to (\ref{Glg1all}) which describes a winding
time dependent texture
for  the case $k=1$ has been found in \cite{TS}. (Time-dependent
solutions to (\ref{Glg1all}) can in fact be found for any $k\in {\bf
N}$ and for any winding number $n\in {\bf N}$ \cite{LukasSpherical}).

The $r$-dependence of the last term 
in (\ref{Glg1all}) shows  that nontrivial solutions of the form 
$\xi_{3}=\xi_{3}(t)$ are impossible for $k\neq 0$. However, for 
static maps the Ansatz (\ref{Ansatz}) becomes 
$\xi_{3}=\xi_{3}(r)$ and equation (\ref{Glg1all}) reduces to 
($\xi\equiv\xi_3$)
\begin{equation}
	\xi^{\prime\prime}+\frac{2}{r}\xi^{\prime}
	-\frac{k(k+1)}{2r^{2}}\sin(2\xi)=0 ~,
	\label{Glg1}
\end{equation}
the equation  we will discuss in the following.

\subsection{Local properties}
\label{ExpansionSect1}
Equation (\ref{Glg1}) has a singular point at $r=0$. Since we require 
solutions to be regular for all $r$ and $t$, we assume that $\xi(r)$ 
is described in a neighborhood of $r=0$ by some power series
\begin{equation}
	\xi(0+\epsilon)=\sum_{j=0}^{\infty}a_{j}\epsilon^{j}.
	\label{PowerSeries}
\end{equation}
If we multiply equation (\ref{Glg1}) by $r^{2}$ and insert 
(\ref{PowerSeries}) at $r=0+\epsilon$ we obtain an equation in powers of 
$\epsilon$
\begin{eqnarray}
    \nonumber
    0 &=& \sum_{j=0}^{\infty}\left((j+2)(j+1)a_{j+2}\epsilon^{j+2}
	+2(j+1)a_{j+1}\epsilon^{j+1}\right)\\
	&-& \frac{k(k+1)}{2}\left[
	\cos(2a_{0})
	\sin\left(2\sum_{j=1}^{\infty}a_{j}\epsilon^{j}\right)
	+\sin(2a_{0})
	\cos\left(2\sum_{j=1}^{\infty}a_{j}\epsilon^{j}\right)
	\right].
\end{eqnarray}

From comparison of the lowest order terms $\epsilon^{0}$ we obtain 
immediately
\begin{equation}
	\sin(2a_{0})=0\; \Rightarrow a_{0}=m\pi/2,\; m\in {\bf Z}.
	\label{Exp1a0}
\end{equation}

Let $l\geq 1$ be the smallest integer for which $a_{l}\neq 0$. Then 
the lowest order terms $\epsilon^{l}$ yield
\begin{equation}
	l(l-1)a_{l}+2la_{l}-\frac{k(k+1)}{2}\cos(2a_{0})\cdot 2a_{l}=0
\end{equation}
Therefore
\begin{equation}
	 a_{l}\left(l(l+1)-k(k+1)\cos(2a_{0})\right)=0.
	\label{PowerResult}
\end{equation}
Combining this with (\ref{Exp1a0}) leads to 
\begin{equation}
	a_l(l(l+1)-(-1)^mk(k+1))=0,
\end{equation}
thus for $m$ odd, $a_l=0$ for all $l>0$ and thus $\xi$ is constant, 
but for $m$ even, $\xi(r)$ has  a 
nontrivial power series expansion at $r=0$ for every eigenmap 
$\Omega:S^{2}\rightarrow S^{2}$ with eigenvalue $k(k+1)$, where the first 
non-vanishing expansion coefficient is just $a_{k}$. For the next higher 
orders $\epsilon^{k+1}$ and $\epsilon^{k+2}$ we get in a similar way
\begin{equation}
	a_{k+1}=0 \mbox{ and } a_{k+2}=-\frac{k(k+1)}{3(2k+3)}a_{k}^{3}.
\end{equation}
Table \ref{KoeffGlg1Tab} summarizes this information.
\begin{table}[tbp]
    \begin{center}
	\begin{tabular}{|c|c|c|}
		\hline
		&&\\
		$a_{0}=$ & $m\pi$ & $(2m+1)\pi/2$  \\
		$a_{j}=~~ (0<j<k)$ & 0 & 0  \\
		$a_{k}=$ & free & 0  \\
		$a_{k+1}=$ & 0 & 0  \\
		$a_{k+2}=$ & $-\frac{k(k+1)}{3(2k+3)}a_{k}^{3}$ & 0  \\
		&&\\
		\hline
	\end{tabular}
    \end{center}
	\caption{Expansion coefficients for $\xi_{3}$ at $r=0$ where 
	$k(k+1)$ is the eigenvalue of the map $\Omega:S^{2}\rightarrow S^{2}$ 
	and $m\in {\bf Z}$.}
	\protect\label{KoeffGlg1Tab}
\end{table}

\subsection{Global Properties}
\label{HomotopySect1}
For a global analysis of the behaviour of solutions of (\ref{Glg1}) 
it is convenient to transform the equation into an autonomous one by 
the  transformation 
\begin{equation}
s=\frac{1}{\beta}\ln r, \mbox{ where }
\beta=\sqrt{\frac{2}{k(k+1)}}.
\label{StaticCoord}
\end{equation}
The new variable $s$ runs from $-\infty$ to $+\infty$. Remember 
that we always require $k>0,\;k\in {\bf N}$, so that $0<\beta\leq 1$.
Denoting the derivative w.r.t. $s$ again by a prime, Eq.~(\ref{Glg1})
transforms into
\begin{equation}
	\xi^{\prime\prime}+\beta\xi^{\prime}-\sin(2\xi)=0
	\label{Glg1e}
\end{equation}

This differential equation describes the 
motion of a particle with constant damping $\beta$ 
and  potential $\sin(2\xi)$. If we switch off the damping, we 
obtain the conservative system 
\begin{equation}
	\xi^{\prime\prime}=-\mbox{grad}U(\xi),
\end{equation}
where the potential $U(\xi)$ is given by
\begin{equation}
	U(\xi)=\int_{\xi_{0}}^{\xi}-\sin(2\xi)d\xi
	=-\sin^{2}\xi-U_{0}.
\end{equation}
The `energy' of this system 
\begin{equation}
	E(\xi,\xi^{\prime})=\frac{1}{2}\xi^{\prime 2}
	-\sin^{2}\xi-U_{0}
	\label{ConsE}
\end{equation}
is conserved, and all solutions are periodic, lying on the sets 
$E$=const. 
When we switch on the damping, the solutions no longer remain on 
levels with $E$=const., but `fall down into the potential wells' at 
$((2m+1)\pi/2,0)$ (see Fig.~1). \vspace{0.4cm}\\
\begin{figure}[htbp]
	\centerline{\psfig{figure=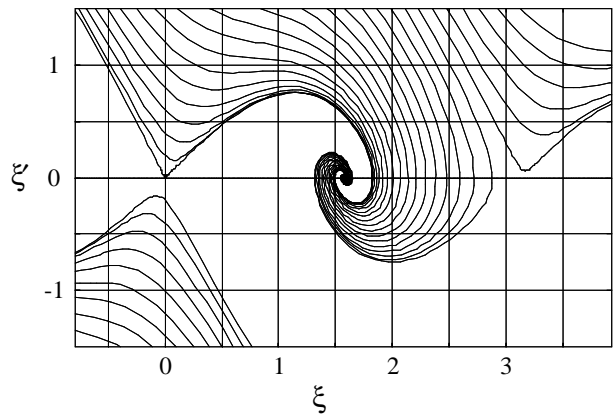,width=8cm}}
	\caption{Phase diagram for the solutions $\xi(s)$ and
$\xi'={d\xi\over ds}$
    of the damped system $(k=1,\beta=1)$.
	Because of the energy loss the solutions
	`fall down' into the potential wells at $((2m+1)\pi/2,0)$}
	\protect\label{StatTextPhaseU}
\end{figure}

If the damping is weak 
($\beta\ll 1$) they spiral several times around those points, 
gradually loosing energy and moving towards the center 
(under-damped motion), whereas if the 
damping is strong ($\beta\gg 1$) they move towards the center quickly
(over-damped motion). In any case, since the energy is no longer 
conserved, non-trivial periodic solutions 
are not possible.

We can formulate this more precisely: If we differentiate the `energy' 
(\ref{ConsE}) with respect to $s$ and insert (\ref{Glg1e}) we obtain
\begin{equation}
	\frac{dE}{ds}=\xi^{\prime}\xi^{\prime\prime}-\sin(2\xi)\xi^{\prime}
	=-\beta\xi^{\prime 2}.
\end{equation}
Thus the energy is always decreasing with growing $s$ 
(for non-constant $\xi$). With this we can show the following Lemma:

\begin{lemma}
\label{Lemma1}
If $\xi(s_{0})=m\pi,\;m\in {\bf Z}$, for some $s_{0}>-\infty$, and if $\xi$ is 
not constant, it tends monotonically to $\pm\infty$ for $s\rightarrow 
-\infty$. 
\end{lemma}

\paragraph{Proof:}
(The idea of the proof is taken from \cite{Bizon}).
Let $\xi(s_{0})=m\pi$ with $\xi^{\prime}(s_{0})=a$. For $s<s_{0}$ we have
\begin{equation}
	0<E(s)-E(s_{0})=\frac{1}{2}\xi^{\prime 2}-\sin^{2}\xi-\frac{1}{2}a^{2}
	\leq \frac{1}{2}\xi^{\prime 2}-\frac{1}{2}a^{2}
\end{equation}
and thus $\xi^{\prime 2}>a^{2},\;\forall s<s_{0}$. Therefore $E$ is 
monotonically increasing for $s\rightarrow -\infty$ and $\xi^{\prime 
2}\rightarrow\infty$ with $\xi^{\prime 2}>0, \; \forall s<s_{0}$. 
Correspondingly, $\xi$ tends monotonically to $\pm\infty$, the sign 
depending on the sign of $\xi'$.
\begin{lemma}
\label{Lemma2}
If $\xi(s_{0})=m\pi,\; m\in {\bf Z}$ for some $s_{0}$, and 
$\xi^{\prime}(s_{0})=0$, then if $\xi$ is not constant either
$m\pi<\xi(s)<(m+1)\pi$ or $(m-1)\pi<\xi(s)<m\pi~~ \forall s>s_0$.
\end{lemma}

\paragraph{Proof:}
Let $\xi(s_{0})=m\pi$ with $\xi^{\prime}(s_{0})=0$. If $\xi$ is not
constant then for $s>s_{0}$ we have 
\begin{equation}
	0>E(s)-E(s_{0})=\frac{1}{2}\xi^{\prime 2}-\sin^{2}\xi
\end{equation}

and thus $\sin^{2}\xi>\frac{1}{2}\xi^{\prime 2}\geq 0,\; \forall s>s_{0}$.
Especially, $\sin^2\xi(s)\neq 0~~ \forall s>s_{0}$ which implies our
statement. 

\begin{corollary}
There is no static texture solution which satisfies the ansatz
(\ref{Ansatz}) ( a texture solution being a solution with homotopy 
degree $\neq 0$, i.e., one that really winds).
\end{corollary}

\paragraph{Proof:}
From Lemma \ref{Lemma1}: The only non-constant regular solutions through a 
point $m\pi$ are the ones with
\begin{equation}
	\xi(-\infty)=\lim_{s\rightarrow -\infty}\xi(s)=m\pi.
\end{equation}
Furthermore, for a regular solution 
\begin{equation}
    \xi^{\prime}(-\infty)= \lim_{s\rightarrow -\infty}\xi^{\prime}(s)=
	\lim_{r\ra 0} \beta r{d\xi\over dr}= 0.
\end{equation} 
To fully wind
around $S^3$, $\xi=\xi_3$ with $\xi=m\pi$ at $r=0 ~~(s=-\infty)$
 would have to  assume either the value $(m-1)\pi$ or $(m+1)\pi$
which, according to Lemma \ref{Lemma2} is not possible. 

\subsection{Existence and stability of static solutions}
Finally, we would like to briefly verify that non-constant solutions 
to (\ref{Glg1e}) do indeed exist globally, and we want to 
review their properties.
With the substitution
\begin{eqnarray}
	x & = & \xi\\
	y & = & \xi^{\prime}
	\label{Substitution}
\end{eqnarray}
equation (\ref{Glg1e}) is equivalent to the autonomous system of first 
order differential equations 
\begin{eqnarray}
	x^{\prime} & = & y\\
	y^{\prime} & = & \sin(2x)-\beta y
\end{eqnarray}
Together with some initial conditions $x(0) = u_{0},\; y(0)=v_{0}$, this 
is an initial value problem (IVP) of the form 
\begin{eqnarray}
    \nonumber
    \left(
	\begin{tabular}{c}
		$x^{\prime}$  \\
		$y^{\prime}$  
	\end{tabular}
	\right)
	& = & f(x,y)\\ 
	(x(0),y(0)) & = & (u_{0},v_{0}).
	\label{Glg1IVP}
\end{eqnarray}
The local existence and uniqueness of a solution to the IVP
(\ref{Glg1IVP}) follow from standard theorems on ordinary differential
equations.

All regular solutions $(x(s),y(s))$ with 
\begin{equation}
    \lim_{s\rightarrow -\infty}x(s)=m\pi,\; m\in{\bf Z} 
    \;(\mbox{and thus }\lim_{s\rightarrow -\infty}y(s)=0)
    \label{StaticBoundary}
\end{equation} 
are bounded for every 
$s\in{\bf R}$ as follows directly from Lemma \ref{Lemma2} and its 
proof. Therefore these solutions exist globally (cf. \cite{ODEs}, 
Corollary 3.2). (The local  existence of  solutions fulfilling 
(\ref{StaticBoundary}) follows from our series expansion in 
Section~\ref{ExpansionSect1}.)

Now let us discuss the stability of the critical points
$(x_{0},y_{0})$ for this system which are given by 
\begin{equation}
	(x_{0},y_{0})=(m\pi/2,0), \; m\in {\bf Z}.
	\label{CriticalPoints}
\end{equation}
In some neighborhood of a critical point we can approximate 
(\ref{Glg1IVP}) by the linearized system
\begin{equation}
    \left(
	\begin{tabular}{c}
		$x^{\prime}$  \\
		$y^{\prime}$  
	\end{tabular}
	\right)
	=(Df)(x_{0},y_{0})
	\left(\begin{tabular}{c}x\\y\end{tabular}\right)
	\label{Glg1linsys}
\end{equation}
where $Df$ is the first derivative of $f$ (with respect to $x$ 
and $y$).
If we calculate the eigenvalues of $Df(x_0,y_0)$ and use the principle of 
linearized stability \cite{Amann} we find that the critical points 
$(x_{0},y_{0})=(m\pi,0),\;m\in {\bf Z}$ are unstable, while the
 critical points 
$(x_{0},y_{0})=((2m+1)\pi/2,0)$,  are stable, 
and attractive (in the sense, that if we start 
at any point close enough to $(x_{0},y_{0})$,  we will always end up 
at the critical point itself for $s\rightarrow\infty$). 
The solutions thus spiral into a focus at $((2m+1)\pi/2,0)$. 
This behavior is clearly seen in Fig.~1: All trajectories 
that come close to the points $(\xi,\xi')=(0,0)$ or
 $(\xi,\xi')=(\pi,0)$ are repelled and
spiral into $\xi=\pm\pi/2$ and $\xi=3\pi/2$ respectively, depending on
the sign of $\xi'$. 

In Fig.~1 we show the phase diagram for $\xi(s)$ and
$\xi'={d\xi\over ds}$ for the case $k=1 (\beta=1)$. Only the solutions with
\[\lim_{s\ra-\infty}\xi(s)=\xi(r=0)=m\pi/2 ~~\mbox{ and}~~~~~ 
\lim_{s\ra-\infty}{d\xi\over ds}=0\]
yield regular solutions in physical space.
Remember that from our power series expansion we need 
$\xi(r=0)=m\pi,\; m\in {\bf Z}$ for a non-constant solution 
(regular at $r=0$). Furthermore, since the `energy' (\ref{ConsE}) is always 
decreasing there are no periodic solutions, and thus all non-constant 
regular solutions 
must end in one of the two focal  points $((2m\pm 1)\pi/2,0)$.
We can therefore conclude: 

\begin{proposition}
The only non-constant solutions to 
equation (\ref{Glg1}) with $k>0$ that are regular for all $r$ are the 
ones starting 
at $\xi(r=0)=m\pi$ and ending in a focus at 
$\xi(r\rightarrow\infty)=(m\pi\pm\pi/2)$ without ever leaving the strip 
$[m\pi,m\pi+c]$ respectively $[m\pi-c,m\pi]$ for $m\in {\bf Z}$ and 
some $0<c<\pi$.
\end{proposition}

This result is also visible in Fig.~1.

\section{Conclusions}
We have found that static 'spherically symmetric' harmonic maps
(solutions of the non-linear $\sigma$-model) from compactified
${\bf R}^3$ into $S^3$ which satisfy our 'ansatz' cannot have
non-trivial topology. Therefore, if a static winding solution exists,
it cannot be represented as the tensor product of a map of the
spherical angles (our $\Omega$) and a radial function. It is not clear
to us, whether such a decomposition should always exist globally. 

We also could not show that each static solution should be homeotopic
to a spherically symmetric static solution and thus spherical symmetry
remains a non-trivial condition which we have to pose.

In this sense, our partial result only hints to the following which
still remains to be fully proven (if true!):

\begin{conjecture}
There exist no static textures with everywhere finite energy density.
\end{conjecture}

\end{document}